\documentclass[12pt]{spieman}  
\usepackage{amsmath,amsfonts,amssymb}
\usepackage{hyperref}
\usepackage{graphicx}
\usepackage{setspace}
\usepackage{tocloft}
\usepackage{url}
\usepackage{textgreek}
\usepackage{bm}
\usepackage{algpseudocode}
\usepackage{algorithm}
\usepackage{amsmath}
\usepackage{lineno}
\usepackage{xcolor}
\usepackage{subfig}
\usepackage{comment}
\DeclareMathOperator*{\argmax}{arg\,max}

\newcommand{\action}{\boldsymbol{a}_t}
\newcommand{\obs}{\boldsymbol{o}_t}
\newcommand{\st}{\boldsymbol{s}_t}
\newcommand{\stp}{\boldsymbol{s}_{t+1}}

\title{Laboratory Experiments of Model-based Reinforcement Learning for Adaptive Optics Control}

\author[a,g]{Jalo Nousiainen}
\author[b]{Byron Engler}
\author[b]{Markus Kasper}
\author[c]{Chang Rajani}
\author[a]{Tapio Helin}
\author[b,e,f]{C{\'e}dric T. Heritier}
\author[d]{Sascha P. Quanz}
\author[d]{and Adrian M. Glauser}

\affil[a]{LUT University, Yliopistonkatu 34, FI-53850, Lappeenranta, Finland}
\affil[b]{European Southern Observatory, Karl-Schwarzschild-Str. 2, 85748, Garching bei M\"{u}nchen, Germany}
\affil[c]{University of Helsinki, Yliopistonkatu 4, FI-00100 Helsinki, Finland}
\affil[d]{ETH Zurich, Institute for Particle Physics \& Astrophysics, Wolfgang-Pauli-Str. 27, 8093 Zurich, Switzerland}
\affil[e]{DOTA, ONERA, F-13661 Salon cedex Air - France}
\affil[f]{Aix Marseille Univ, CNRS, CNES, LAM, Marseille, France}
\affil[g]{Aalto University, Department of Mathematics and Systems Analysis, P.O. Box 11100, FI-00076 Aalto, Finland}

\cftpagenumbersoff{figure}
\cftpagenumbersoff{table} 
\begin{document} 
\maketitle
\begin{abstract}
Direct imaging of Earth-like exoplanets is one of the most prominent scientific drivers of the next generation of ground-based telescopes. Typically, Earth-like exoplanets are located at small angular separations from their host stars, making their detection difficult. Consequently, the adaptive optics (AO) system's control algorithm must be carefully designed to distinguish the exoplanet from the residual light produced by the host star.

A new promising avenue of research to improve AO control builds on data-driven control methods such as Reinforcement Learning (RL). RL is an active branch of the machine learning research field, where control of a system is learned through interaction with the environment. Thus, RL can be seen as an automated approach to AO control, where its usage is entirely a turnkey operation. In particular, model-based reinforcement learning (MBRL) has been shown to cope with both temporal and misregistration errors. Similarly, it has been demonstrated to adapt to non-linear wavefront sensing while being efficient in training and execution.

In this work, we implement and adapt an RL method called Policy Optimization for AO (PO4AO) to the GHOST test bench at ESO headquarters, where we demonstrate a strong performance of the method in a laboratory environment. Our implementation allows the training to be performed parallel to inference, which is crucial for on-sky operation. In particular, we study the predictive and self-calibrating aspects of the method. The new implementation on GHOST running PyTorch introduces only around 700 \textmu s of in addition to hardware, pipeline, and Python interface latency. We open-source well-documented code for the implementation and specify the requirements for the RTC pipeline. We also discuss the important hyperparameters of the method and how they affect the method. Further, the paper discusses the source of the latency and the possible paths for a lower latency implementation. 
\end{abstract}


{\noindent \footnotesize\textbf{*}Jalo Nousiainen,  \linkable{jalo.nousiainen@aalto.fi} }

\begin{spacing}{2}   

\section{Introduction}
\label{sect:intro}  
High contrast imaging (HCI) utilizes a combination of eXtreme Adaptive Optics (XAO) and coronagraphy to generate images of faint sources near bright point sources, such as exoplanets near their host stars. Direct imaging of exoplanets has been largely limited to only a few dozen very young and luminous giant exoplanets using existing HCI instruments, as documented in studies such as \cite{marois2010images, lagrange2009probable, macintosh2015discovery}. However, a greater number of planets could be directly imaged by enhancing the sensitivity in the vicinity of the host star, with the performance of the XAO system being the primary limiting factor in achieving such sensitivity.

In HCI, when imaging in close proximity to the star, the main performance limitations of a well-tuned adaptive optics (AO) system controlled with the common integrator controller are photon noise and temporal error, as noted in \cite{guyon2005limits}. The temporal delay error of AO systems controlled by standard methods arises from the integration of wavefront sensor detector data, detector readout, computation of the correction signal, and its application to the deformable mirror (DM). This delay amounts to at least two AO system operating cycles at the maximum camera framerate, where readout takes one entire frame, during which atmospheric turbulence has evolved and no longer matches the DM correction precisely.

There are two ways to mitigate the adverse effect of the temporal delay error for HCI: by increasing the operating frequency of the AO system or by implementing predictive control. The acceleration of the AO system can be accomplished, for example, by adding a second stage downstream from a classical first-stage AO system \cite{cerpa2022cascade}. This second-stage system solely observes the residual from the first-stage AO system and can operate independently from the first-stage, employing DMs that can handle fast AO loops. One such example is the upgrade of SPHERE, which is referred to as SPHERE+ \cite{boccaletti2020sphere+},  which is expected to provide a considerable enhancement in raw point-spread function (PSF) contrast close to the star.

The other (not mutually exclusive) approach is to use a predictive control algorithm. A big part of the turbulence is presumably in frozen flow considering the millisecond timescale of AO control, and hence, a significant fraction of wavefront disturbances can be predicted \cite{poyneer2009experimental}. Moreover, if the predictive control algorithm is fast enough, both strategies can be combined by operating the faster second stage with predictive control.


Besides the performance limitations induced by photon noise and temporal error, AO can suffer from dynamic modeling errors such as misregistration \cite{heritier2018new}, optical gain effect for the Pyramid WFS \cite{korkiakoski2008improving, deo2019telescope}. Coping with these limitations usually requires external tuning and recalibration of a possible predictive control algorithm.

This paper presents a laboratory demonstration of a data-driven predictive control algorithm called the Policy Optimizations for AO (PO4AO) \cite{nousiainen2022toward} implemented on a second stage AO system following a first stage running a classical integrator control. One of the main advantages of implementing fully data-driven control, such as PO4AO, is that it continuously learns a system model from the data rather than using a static calibration or synthetic model. Consequently, it is less affected by pseudo-open-loop reconstruction errors, such as misregistration or the optical gain effect, as discussed in \cite{nousiainen2021adaptive, haffert2021data, nousiainen2022toward}. Our contributions are two-fold: first, we thoroughly test the performance and robustness of PO4AO in a laboratory setup, detailed in Section \ref{sec:ghost}, under different conditions. Second, we open-source Python-based implementation of the method that can be implemented in any AO system that runs Python-based controllers and has GPUs. We also discuss how the method can be tuned and further developed for different AO systems. The codes used in this paper are available on our GitHub repository [\href{https://github.com/jnousi/PO4AO.git}{https://github.com/jnousi/PO4AO.git}].

\section{Related work}
PO4AO addresses the predictive control and reconstruction in the XAO control loop as a single reinforcement learning problem; hence, PO4AO is related to many aspects of XAO control, such as predictive control, optimal gain compensation, misregistration identification, reconstruction algorithms, and vibration canceling.

Remarkable progress has been achieved with various approaches to tackle the XAO control problem. These methods include the Kalman filter-based linear controllers \cite{kulcsar2006optimal, paschall1993linear, gray2012ensemble, conan1a2011integral, correia2010adapting, correia2010optimal, correia2017modeling}, sometimes combined with machine learning for system identification \cite{sinquin2020sky}. These methods rely on linear models for wavefront sensing and temporal evolution to obtain a state estimation of the system. Other methods focus on correcting temporal error and vary from spatio-temporal linear filters to filters operating on single modes, such as Fourier or Zernike modes \cite{guyon2017adaptive, poyneer2007fourier, dessenne1998optimization, van2017performance, van2019impact, males2018ground}. The predictive filters are obtained either from modeling or utilizing data analysis/machine learning. Some methods have also been tested on-sky (see, e.g., \cite{van2022predictive, sinquin2020sky}).

Moreover, machine learning methods utilizing neural networks (NN) for predictive control have been studied in \cite{swanson2018wavefront, sun2017bayesian, liu2019using, wong2021predictive, swanson2021closed, hafeez2022forecasting}, where NNs show a lot potential, especially for AO systems with high number of degrees of freedom (DoF), and in noisy conditions. Lately, NNs, and the more modern deep NNs, have also been used for the wavefront reconstruction step (see, e.g., \cite{Landman:20, wong2023nonlinear, archinuk2023mitigating, he2021deep}). The results indicate that NN reconstruction is less sensitive to non-linearity and increases the operational range of the Pyramid WFS. 

Lastly, different NN-based reinforcement learning approaches have been studied during the last years; see \cite{Pou:22, landman2020self, landman2021self}. PO4AO differs from other RL methods in AO literature by using so-called model-based RL instead of model-free RL (for a discussion on the difference between these methods, see \cite{nousiainen2022toward}). For interested readers,  Fowler \& Landman \cite{fowler2023tempestas} provide a more thorough review of machine learning methods for wavefront control and phase prediction.

\section{Classical adaptive optics control and baseline controller}
An AO system is commonly controlled with a linear integrator controller, referred to as the integrator. We consider it our reference method against the PO4AO as it is still widely used in AO. Integrator control in AO usually relies on the so-called interaction matrix mapping DM commands to WFS measurements:
 \begin{equation}
    \Delta \bm w^t = D \bm v^t + \xi_t,   
\end{equation}
where $\Delta \bm w^t = (\delta w_1^t,\delta w_2^t, \cdots, \delta w_n^t)$ is the WFS data, $\bm v^t$ the DM commands and $D$ is the interaction matrix and $\xi_t$ is the measurement noise typically composed of photon and detector noise.
Once the interaction matrix is estimated, the inverse problem, i.e., reconstruction $\bm v^t$ given $\Delta \bm w^t$, needs to be considered. As $D$ is generally not invertible, some regularization approach is needed. Here, we restrict ourselves to linear methods described by a reconstruction matrix $C$ mapping WFS measurements to DM commands. As our regularization method, we project $D$ to a smaller dimensional subspace spanned by the Karhunen--Lo\'{e}ve (KL) modal basis \cite{1994ESOC...48..187G}. Each KL mode in the basis has a representation in terms of actuator voltages. This relation is fully determined by a transformation matrix $B_m$ mapping DM actuator voltages to $m$ first modal coefficients. The regularized reconstruction matrix is now defined by the Moore--Penrose pseudo-inverse
\begin{equation}
C_m = (D P_m)^{\dagger},
\end{equation}
where $P_m = B_m^\dagger B_m$ is a projection map to the KL basis.
The number of modes $m$ defines stability at the cost of resolution; smaller $m$ results in lower noise amplification while producing a reconstruction with fewer modal basis functions (less detailed reconstruction). An optimal $m$ balances the error produced by these two effects.

We use the leaky integrator as a baseline AO controller to which PO4AO is compared. At a given time step $t$, the WFS measures the residual wavefront. The leaky integrator then obtains the new control voltages $\Tilde{\bm v}_t$ from
\begin{equation}
\label{eq:lin_integrator}
\Tilde{\bm v}_t = l\Tilde{\bm v}_{t-1} + g C_m \Delta \bm w_t,
\end{equation}
where $g$ is the integrator gain, typically fixed below a value of about 0.5 for a two-step delay system \cite{madec1999control}. The
DM saturation can cause a build-up of modes outside the control radius. Hence, introducing a leakage $l$ typically chosen near one, e.g., 0.99, in the DM commands commonly used to remove those unseen modes and increase robustness.

\section{Adaptive optics as a Markov decision process}\label{sec:mdp}
In this paper, we model AO control loop as a Markov decision process (MDP) \cite{bellman1957markovian}, which is the de facto mathematical framework for sequential decision problems in RL. In AO control, instead of the full state of the system (exact DM shape, the full atmosphere profile, etc.), we only observe WFS data $\Delta \bm w_t$ that represents only a partial information of the whole system \cite{nousiainen2021adaptive}. These kinds of processes are referred to as Partially observed MDPs in the RL literature, and, in theory, optimal decisions (control) should consider all the past measurements observed (from the beginning of operation). However, we expand the state space to include a history of WFS measurement and DM control voltages to guarantee approximately Markovian statistics and treat the process as an ordinary MDP, where optimal decisions can be taken directly from the previous state formulation (in our case a concatenation of past measurement and actions) \cite{nousiainen2021adaptive, nousiainen2022toward, Pou:22, landman2021self}.

Let us first identify an action $\action$ as the applied differential voltage satisfying 
\begin{equation}
 \Tilde{\bm v}_t = l  \tilde{\bm v}_{t-1} + \action,
\end{equation}
where $l$ is the leakage factor.

We now define the state $\bm s_t$ of the MDP as
\begin{equation}
\bm s_t =  \begin{pmatrix} \bm o_{t}, \bm o_{t-1}, \dots, \bm o_{t-k}, \bm a_{t-1}, \bm a_{t-2}, \dots, \bm a_{t-k} \end{pmatrix},
\end{equation}
where $\obs = C_m \Delta \bm w_t$ is the wavefront measurement projected to DM space and $k$ the number of history frames used in state formulation. Hence, we assume that there exist Markovian transition dynamics $p(\bm s_t, \bm a_t) = \bm s_{t+1}$, where the only new element of $\bm s_{t+1}$ is the next observation $\bm o_{t+1}$. This transition model contains information on the time delay (i.e., which action affects which observation), misregistration and non-linearity errors (how actions are related to measurements), and atmospheric turbulence (how past information can be interpolated in the future). Further, in the following formulation of the control Algorithm \ref{sec:algo}, the initial reconstruction matrix $C_m$ serves as preprocessing to observation and does not connect measurement to action directly (like in integrator control).

As the reward function of the MDP, we consider the negative Euclidian norm of the residual wavefront observed through the WFS with a regularization term that favors small actions, i.e.,
\begin{equation}\label{eq:reward}
    \hat r_\omega(\st, \bm s_{t+1}, \action) = - \|\bm o_{t+1}\|^2 - \alpha \|\action\|^2,
\end{equation}
where $\bm o_{t+1}$ is the first element of $\bm s_{t+1}$. The addition of the regularization term $ \alpha \|\action\|^2$ effectively regularizes the control algorithm, making it less prone to saturation and oscillation, especially early in the training procedure.

\section{Model-based Policy optimization}\label{sec:algo}
The key idea of PO4AO (for details, see Nousiainen et al. 2022 \cite{nousiainen2022toward}) is to learn a non-linear control law that maps past telemetry to new DM commands from data collected from the AO loop and maximizes the reward. In RL terminology, this control law is referred to as the \emph{policy} and will be formulated as a mapping from the current state $\bm s_t$ to the next action $\bm a_t$. Hence, the policy combines the reconstruction and control steps in AO (e.g., a least-squares modal reconstruction followed by integrator control).

\begin{figure}
\begin{center}
\begin{tabular}{c}
\includegraphics[height=6.7cm]{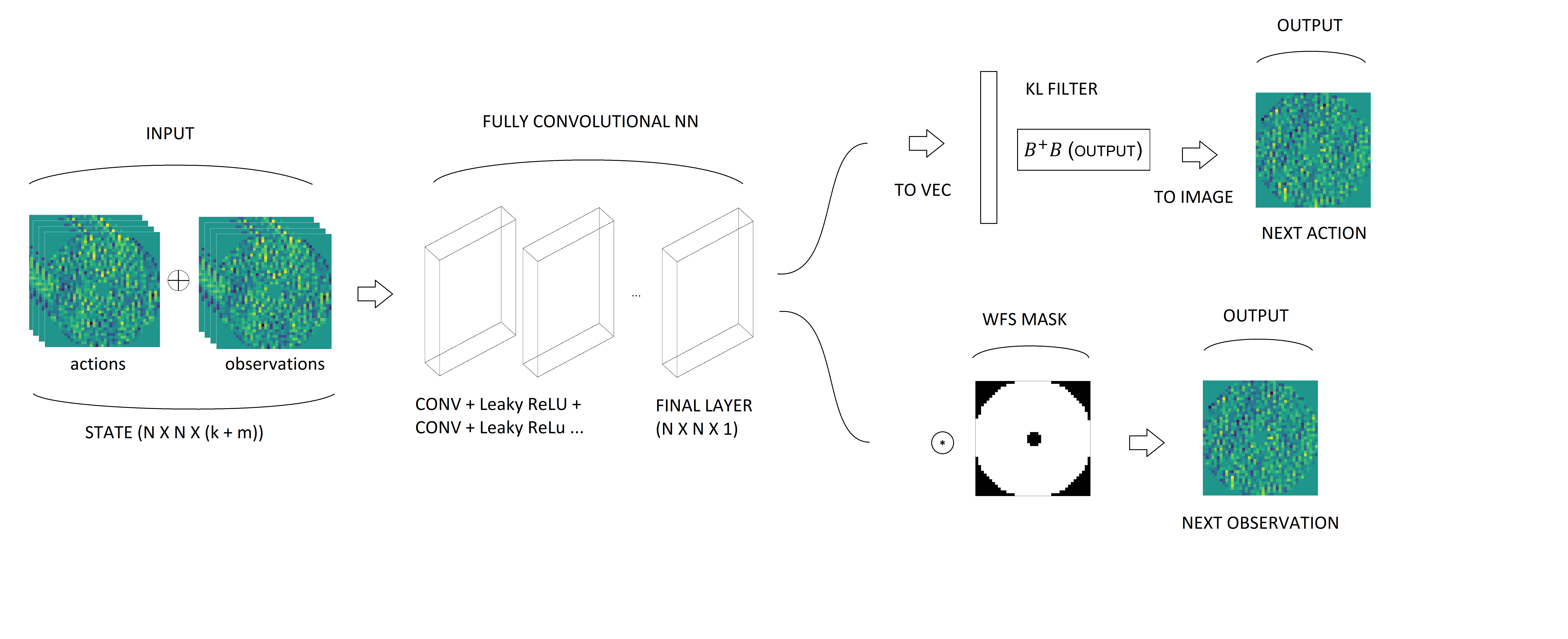}
\end{tabular}
\end{center}
\caption 
{ \label{fig:nnmodels}
Neural network architectures. Both NN, the dynamics, and policy take input tensor concatenations of past actions and observations. They also share the same fully convolutional structure in the first 3-layers. At the output layer, the policy model includes the KL-filtering scheme (upper right corner), and the dynamics model output is multiplied with the WFS mask (lower right corner). For the GHOST, the input and output images are 24x24 pixels (set by the DM).}
\end{figure}

In this work, the policy is constructed as a neural network, and its parameters are derived indirectly via model-based policy optimization. More precisely, the method collects data to learn a dynamics model that is also represented by an NN and can be used to predict the subsequent state given the current state and an action. The dynamics model is then used to optimize the policy. Both neural networks are Fully Convolutional NN with three layers, see Figure \ref{fig:nnmodels}

The method first runs the so-called warm-up phase, where an initial data set is collected by injecting random control signals into the control system, followed by training the involved NN models. The warm-up phase aims to ensure rough estimates of policy and dynamic NN and, consequently, stabilize the training procedure in the beginning. After the warm-up phase, the method iterates the following three phases, from which phase one (1) is run in parallel to phases two (2) and three (3).  
\begin{enumerate}
    \item \textbf{Running the policy:} The method collects data by running the policy in the AO control loop for $T$ timesteps (a single episode).
    \item  \textbf{Improving the dynamics model:} The dynamics model parameters are optimized via a supervised learning objective.
    \item  \textbf{Improving the policy:} The policy parameters are optimized by utilizing the dynamics model.
\end{enumerate}
Let us now describe the PO4AO algorithm in more detail. The dynamics model $\hat{p}_{\omega} : (\st, \action)\mapsto \bm s_{t+1}$ parametrized by $\omega$ is expressed as an ensemble, i.e., a collection of deterministic convolutional neural networks (CNNs), where $\omega$ represents the weights and biases of the networks. Moreover, the policy mapping $\pi_\theta : \st \mapsto \bm a_{t}$ parametrized by $\theta$ is constructed as a fully CNN followed by a modal filter layer, i.e., 
\begin{equation}
\pi_\theta(\st) =  P_m F_\theta(\st),    
\end{equation}
where $P_m$ is the projection map, and $F_\theta$ is a fully CNN, where the output is vectorized, and $\theta$ represents the weights and biases of the CNN.

In {\bf Step 1}, telemetry (tuples of $(\st, \action, \stp)$ saved into a dataset $\mathcal{D}$) is collected by operating the AO control loop with the current (or initial) parametrization of the policy. In practice, PO4AO utilizes two datasets: one for the data collected during the warm-up and one for the most recent data (see details in section \ref{sec:imp}).

Utilizing this data, in {\bf Step 2}, the dynamics model is trained, i.e., the parametrization is optimized by minimizing the squared difference between the true next states and the predictions according to
\begin{equation}
\label{eq:dynamics_error}
    \sum_{\mathcal{D}} \left\|\stp - \hat{p}_\omega(\st, \action)\right\|^2
    = \sum_{\mathcal{D}} \left\|\bm o_{t+1} - \hat{\bm o}_{t+1}\right\|^2,
\end{equation}
where ${\bm o}_{t+1}$ is obtained from the state $\stp$ and $\hat{\bm o}_{t+1}$ is the observation predicted by $\hat{p}_\omega(\st, \action)$. The parameters are optimized by the Adam algorithm \cite{kingma2014adam}. 

The objective of {\bf Step 3} is to find policy parameters $\theta$ that maximize the expected reward within some pre-defined time horizon $H$ given the dynamics of the environment (in our case, the approximate model $\hat{p}_\omega$), that is 
\begin{equation}
    \label{eq:optimization_of_hatr}
    \argmax_\theta \sum_{{\bf s} \in \mathcal{D}}  \sum_{t=1}^{H} \hat r_\omega(\tilde \st, \pi_\theta(\tilde \st)),
\end{equation}
where $H$ is so-called \emph{planning horizon} and
\begin{equation*}
    \tilde {\bf s}_1 = {\bf s} \quad \text{and} \quad \tilde {\bf s}_{t+1} = \hat p_\omega(\tilde \st, \pi_\theta(\tilde \st)).
\end{equation*}
In practice, this is done by sampling from previously observed data points, computing the actions, and using the dynamics model to simulate the future. Moreover, we use the differentiability of the reward and backpropagate through the models. 

We give separate pseudo codes for the warm-up phase and the two parallel processes, which are step one (1), i.e., running the policy, and steps two (2) and three (2), i.e., improving the dynamics model and improving the policy. Algorithms 1, 2, and 3 give a full pseudo-code for the procedures.

\begin{algorithm}
\caption{PO4AO warm up}
\label{alg:warmup}
\begin{algorithmic}[1]
\State Initialize policy and dynamics model parameters $\theta$ and $\omega$ randomly
\State Initialize gradient iteration length $K$, batch size $B<|\mathcal{D}|$ and planning horizon $H$
\For{number of warm-up episodes}
    \For{frame $t$ to $T$}
        \State control with noisy integrator, i.e., $\action = g\Delta \bm v_t +  \sigma x$, where $x$ is Gaussian random noise and full command $ \bm v_t = l\bm v_{t-1} + \action$.  Record data $\{s_{t+1}, s_t, a_t \}$ and append to $\mathcal{D}$
\EndFor
\State reduce $\sigma$
\EndFor
\State Fit dynamics by minimizing Eq. \eqref{eq:dynamics_error} w.r.t $\omega$ using Adam
\State Fit policy by maximizing Eq. \eqref{eq:optimization_of_hatr}
\State Start parallel processes described in Algorithms 2 and 3.
\end{algorithmic}
\end{algorithm}

\begin{algorithm}
\caption{PO4AO control thread}
\label{alg:control}
\begin{algorithmic}[1]
\While{observing}
    \For{frame $t$ to $T$}
        \State wait for camera readout 
        \State run policy $a_t = \pi_\theta(a_t|s_t)$, and send $a_t$ to DM   
        \State record data $\{s_{t+1}, s_t, a_t \}$ append to $\mathcal{D}$
    \EndFor    
\EndWhile
\end{algorithmic}
\end{algorithm}

\begin{algorithm}
\caption{PO4AO training thread}
\label{alg:training}
\begin{algorithmic}[1]
\State Initialize gradient iteration length $K$, batch size $B<|\mathcal{D}|$ and planning horizon $H$
    \While{observing}
  \State Fit dynamics by minimizing Eq. \eqref{eq:dynamics_error} w.r.t $\omega$ using Adam
  \For{iteration $k=1$ to $K$}
      \State Sample a mini-batch of $B < |\mathcal{D}|$ states $\{ s_\tau \}$ from $\mathcal{D}$
      \For{each ${\bf s}_\tau$ in the mini batch}
        \State Set $\tilde {\bf s}_1^\tau = {\bf s}_\tau$
        \For{$t=1$ to $H$}
            \State Predict $\action = \pi_\theta(\st)$
            \State Predict $\stp = \hat{p}_\omega(\st, \action)$
            \State Calculate $R_t = \hat r_\omega(\st, \action)$
        \EndFor
      \EndFor
      \State Update $\theta$ by taking a gradient step according to $\nabla_\theta \sum_{t=\tau}^{ \tau + H} R_t $ with Adam.
      \EndFor
      \EndWhile
\end{algorithmic}
\end{algorithm}

\begin{table*}[ht]
    \centering
    \caption{Table of adjustable PO4AO parameters. The values here were used during the experiments in the results section, and they are a good starting point for tuning the method into new instruments.}
    \label{table:po4ao_param}
\begin{tabular}{ |c| c| c|} 
 \hline
 \multicolumn{3}{|c|}{Reinforcement Learning parameters} \\
 \hline
         Parameter  & Value  &  Units  \\
 \hline
 Episode length   &  500  & frames     \\
 Warm-up episodes    &  20  & episodes   \\
 Initial warm-up noise   &  2  & \% of the maximum stroke \\
 Minimum warm-up noise  &   1   &    \% of the maximum stroke  \\ 
 Reward penalty ($\alpha$) &  0.1 &  $\cdots$  \\ 
 \hline
 \multicolumn{3}{|c|}{Training parameters} \\
 \hline
         Parameter  & Value  &  Units  \\
 \hline  
 Iterations after warm-up dynamics &  300  & Gradient steps        \\
 Iterations after warm-up policy (K) & 150 & Gradient steps \\
 Iterations during episode dynamics & 30 & Gradient steps  \\
 Iterations during episode policy (K) & 15 & Gradient steps  \\
 Mini Batch size in training (B) & 32 & Data samples \\
 \hline
 \multicolumn{3}{|c|}{Markov decision process parameters} \\
 \hline
 Number of history frames     &  32  & frames     \\
 Planning horizon    &  4  & frames       \\
  \hline
 \multicolumn{3}{|c|}{Replay buffers} \\
 \hline
 Replay buffer size    &  20  & episodes     \\
 Warm-up buffer size     &  20  & episodes    \\
 Train warm-up percent        & 20   &   percent (\%)          \\   
  \hline
\end{tabular}
\end{table*}

\section{PO4AO implementation and hyperparameters}\label{sec:imp}
PO4AO has a lot of free adjustable parameters; see Table \ref{table:po4ao_param}. The subsections below discuss the specific choices and how they affect the method's performance. We arrange hyperparameters under four different subcategories.

\subsection{Reinforcement Learning parameters}
These parameters set the frequency on which the policy NN is updated. The episode length ($T$ in the pseudo codes) is the number of frames in an episode; for example, 500 frames on GHOST running at 350 Hz is ~1.4sec. A single training procedure is run during the episode (i.e., the dynamics and the policy optimization). After each episode, the policy is updated with the newly updated model. The episode length determines the maximum speed at which PO4AO can adapt to changing conditions.

The warm-up length determines how many episodes are run in the warm-up phase. For example, if the warm-up length is 20, the first 20 episodes are run with the noisy integrator. The Initial warm-up noise sets the maximum noise variance for the added noise component, and the Minimum warm-up noise sets the minimum. The noise is reduced linearly during the warm-up -- starting from maximum and finishing at minimum. The training procedure is started after the warm-up phase. The noise levels should be adjusted considering the dynamic range of the DM and WFS.  Initially, the control should be really noisy, but we should not saturate the mirror too much. We use 1\% of the full range of the DM. The control should be close to the integrator performance level in the latest warm-up episodes. The progressive reduction of the injected noise amplitudes in the warmup data covers a wide range of actions and states of the system, which is crucial knowledge for PO4AO to avoid instabilities when occasionally facing large wave-front residuals.


The loss function penalty parameter $\alpha$ defines the amount of regularization in the reward function Equation \eqref{eq:reward}. The default value $\alpha = 0.1$ provided enough regularization without affecting the performance.

\subsection{Training parameters}
These parameters set the number of gradient steps in dynamics and policy optimization. After the warm-up phase, the loop is suspended, and the first training procedure is run on the data obtained for the warm-up. The parameter \emph{Iterations after warm-up} sets the number of gradient steps/iterations in this first training procedure. The suspension of the loop could be avoided by doing the warm-up training in parallel to warm-up control. However, suspension offers more flexible implementation for testing required warm-up training time, and the warm-up phase is usually avoided by using a pre-trained model. 

After the first training iteration, the loop is closed with policy, and the parallel training procedure is started. Parameters \emph{Iterations during the episode dynamics and - policy} parameters set the number of gradient steps on the parallel training thread. The latter parameters should be set so that the training procedure finishes during a single episode; for example, with an episode length of 1.4 sec, the training procedure should take a maximum of 1.4 sec. A good rule is to train the dynamics model more than the policy.

The \emph{Mini batch size} is the number of data points used to calculate a single gradient step. These data points are randomly sampled from the dataset (i.e., Replay buffer or warm-up buffer).

\subsection{Markov decision process parameters}
The MDP formulation outlined in Section \ref{sec:mdp} is specified by the parameters under the category \emph{Markov decision process parameters}. The number of history frames, $k$, decides the number of past measurements in the MDP formulation -- The policy (controller) does not remember events outside this horizon. Consequently, periodic disturbances occurring at a rate longer than the history horizon are hard to predict.

Compared to a short history, the long history horizon has advantages. First, it considers more measurements and can effectively average the measurement noise. Secondly, a long history enables the method to predict low-order vibration. On the other hand, longer history increases the computational burden and hampers the training (the bigger the input, the more free parameters to train).

The choice of the number of history frames is not trivial and should be tuned according to the properties of the instrument. In Section \ref{sec:horizon}, we run an experiment with different history lengths and compare the results.

The planning horizon, $H$, sets the future time window considered by the PO4AO (see Section \ref{sec:algo}). The loop latency drives the choice of this parameter, that is, the time delay. A good and stable choice of these parameters is typically 1-2 frames longer than the expected time delay of the system (see detailed discussion in \cite{nousiainen2022toward}).

\subsection{Replay buffers}
PO4AO includes two buffers for saving data: the warm-up and replay buffer. The warm-up buffer saves the data recorded during the warm-up phase and stays the same after the warm-up phase. The warm-up buffer size is usually set to the length of the warm-up phase, that is, in our case, 20 episodes (10k frames). The newest data is added to the replay buffer, which keeps the latest \emph{Replay buffer size} episodes in memory. The mini-batch sampled during the training is sampled from the warm-up buffer with the probability set by the \emph{Train warm-up percentage} and otherwise from the replay buffer. The warmup buffer data with the relatively large Gaussian noise added to the actions are needed to remind the policy of "bad" actions. Suppose the models are only trained with recent data. In that case, PO4AO will eventually forget about the actions outside the good control regime and, consequently, start to explore the region's bad actions again, leading to bad performance. We note here that the data in the bad region does not need to be very accurate -- the PO4AO only has to remember that the good rewards are observed when the loop is behaving well. 

The length of the replay buffer also affects the method's ability to adapt to changing conditions. A short replay buffer will react to changes in condition quickly but is prone to overfitting as the method trains on shorter data sequences. For our system and testing, a replay buffer length of 20 episodes was a good compromise between the two effects.

\section{Experiments}
Here, we present the results of several experiments performed with PO4AO on GHOST to explore its performance for high-level conditions relevant to operational on-sky AO. Specifically, we explore
\begin{enumerate}
    \item The impact of the temporal delay on the performance by adding artificial extra delay (unknown to PO4AO) to the control loop.
    \item The robustness and performance for low S/N
    \item The ability to cope with misregistration 
    \item  The effect of the history length (\emph{number of history frames} in Markov decision process parameters, Table \ref{table:po4ao_param}) on performance. 
\end{enumerate}

In all experiments, the PO4AO is compared against an integrator whose gain is adjusted to minimize WFS residuals in all cases separately. Most other parameters listed in Table \ref{table:po4ao_param} (RL, Training, replay buffers, and NN models) were kept constant. First, we introduce our lab setup in sub-sections \ref{sec:ghost} and \ref{sec:numeric}, then the individual experiments.

\subsection{The GPU-based High-order adaptive OpticS Testbench}\label{sec:ghost}
The GPU-based High-order adaptive OpticS Testbench (GHOST) laboratory AO system \cite{engler2022gpu} has been built to evaluate new AO control techniques, specifically predictive control, for the ELT Planetary Camera and Spectrograph (PCS). Located at the ESO headquarters in Garching, Germany, the GHOST utilizes a simple single-source (single-mode fiber-coupled 770 nm SLED) on-axis setup equipped with a pyramid WFS and a Boston Micromachines (BMC) deformable mirror (DM-492). A programmable Spatial Light Modulator (SLM, Meadowlark HSP1920-600-1300-HSP8) introduces turbulence with high spatial resolution. 

\begin{figure}
\begin{center}
\begin{tabular}{c}
\includegraphics[height=7.5cm]{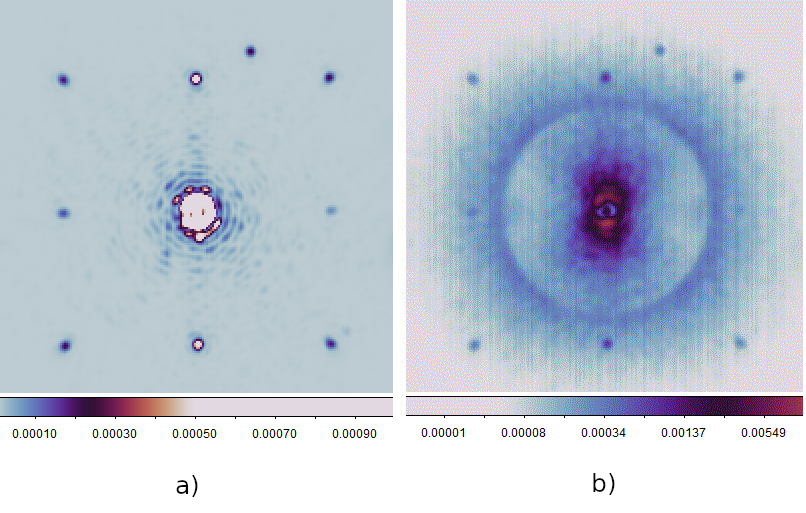}
\end{tabular}
\end{center}
\caption 
{ \label{fig:ghost} GHOST coronagraphic PSFs. Left: the PSF without any turbulence, and the DM set to be flat. Right: the PSF with simulated 1-stage systems residual phase screens played on SLM, and a flat DM. The speckle at around 1 o'clock is a ghost in the system.}
\end{figure}

The GHOST also splits the beam before the WFS to provide a "science channel" with a classical Lyot coronagraph (4 \textlambda/D diameter mask and a circular Lyot stop undersized to 85\%) and a CMOS camera Basler ACA2040-90um) with a sampling of $\sim 3 \text{pixel} / \lambda /\text{D}$. Figure \ref{fig:ghost} shows the coronagraphic PSF limited by the bench aberrations (left), and the long-exposure coronagraphic PSF during replay of the numerically simulated 1st stage residuals by the SLM (right). The grid of satellite spots is produced by the actuator structure of the DM-492. The brightest of these spots is about 8e-4 of the intensity of the central PSF with the Lyot mask removed.

The Real-time computer (RTC) is built using Commercial off-the-shelf (COTS) server components and two Nvidia RTX Titan Graphics processing units (GPU)s. As the software solution, the GHOST makes use of the COSMIC RTC  \cite{ferreira2020hard}. The COSMIC RTC is a platform developed for adaptive optics real-time control (AO RTC) and proposed for several future AO instruments. 



The GHOST simulates a two-stage XAO system closely resembling the VLT/SPHERE+ setup. The initial control phase is conducted via simulation, utilizing a pyramid or SH-WFS (40 x 40 grid), effectively overseeing 800 modes. The atmosphere in the numeric simulation is sampled at twice the rate of the control loop. The residual wavefront error is subsequently captured and applied to the setup via an SLM. The second stage control utilizes the actual hardware, a pyramid wavefront sensor coupled with the DM controlling 300-400 modes, at the SLM frequency (the second stage runs effectively twice the first stage frequency). We operate the SLM and DM at the maximum (stable) frequency of the SLM, 350 Hz, in all experiments.


\begin{figure}
\begin{center}
\begin{tabular}{c}
\includegraphics[height=5.7cm]{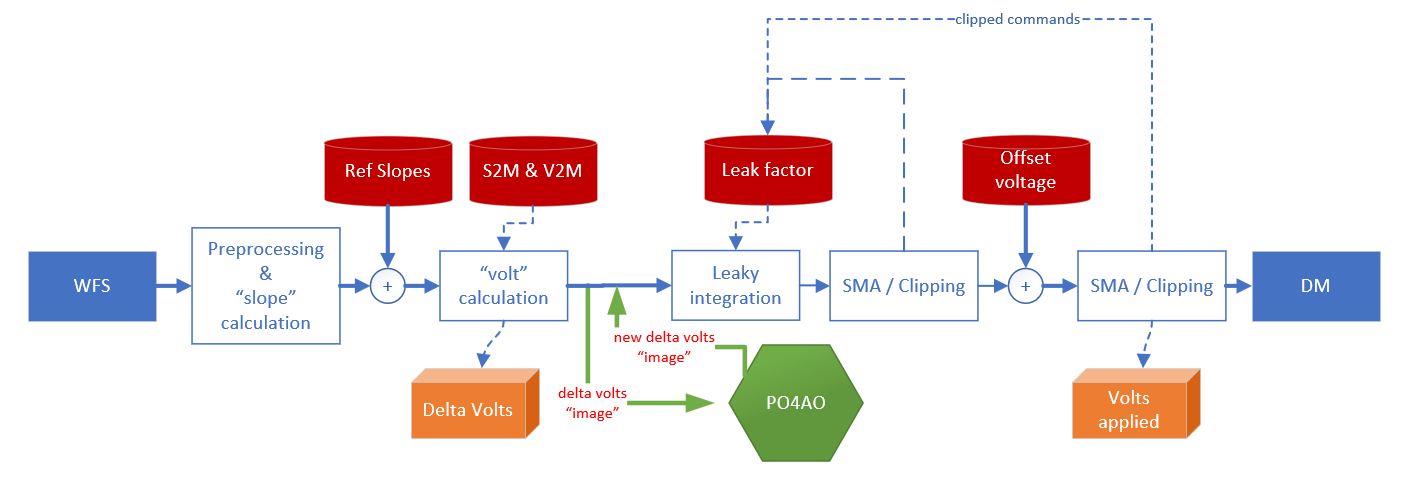}
\end{tabular}
\end{center}
\caption 
{ \label{fig:rtc}
PO4AO interface for RTC pipeline. COSMIC pipeline preprocesses the raw WFS data, projects it to DM-space with command matrix (using the modal basis matrices: S2M and V2M), then writes the "delta volts" to the shared memory buffer, and suspends the loop. Python interface (the green box) reads the shared memory buffer and passes the data to the PO4AO implementation. The PO4AO calculates the next command and saves the data (orange boxes), and the Python interface  writes the command to shared memory, where COSMIC registers the command and passes it to the saturation management algorithm (SMA) / clipping stage.} 
\end{figure}

\subsubsection*{Implementation to GHOST and interfacing to COSMIC pipeline}
The Python script that runs the PO4AO is composed of two threads. One thread (called the control thread) controls the system with the trained policy NN by reading and writing to the shared memory buffers. The other thread (called the training thread) constantly trains the models, i.e., optimizes the dynamics and policy model parameters by running lines 6-17 in the Algorithm \ref{alg:training}. Once a single episode is run, the newly updated policy parameters are imported to the control thread.

The PO4AO algorithm interfaces to the COSMIC pipeline through a shared memory buffer. The COSMIC pipeline reads the raw WFS images, preprocesses them to slopes, and adds the reference slopes; then, it multiplies the slopes with the reconstruction matrix, writes the resulting delta voltages to the shared memory buffer, and suspends the loop. The Python script reads the delta voltages from the shared memory and passes the data to PO4AO. The PO4AO keeps past residual voltages and actions applied in memory. The memory of past delta voltages and actions and the new delta voltages are passed to policy NN that outputs a two-dimensional image of voltages, the action, which is then written to DM command buffer where COSMIC catches them. The control thread also saves the observed delta voltages and actions fed (the orange boxes in the figure) to the replay buffers for the training.

\subsection{Simulation set-up}\label{sec:numeric}
To simulate a faster second-stage system using GHOST, we followed these steps. Firstly, we generated the residual phase screens numerically for the lab setup by using the Object-Oriented Python and Adaptive Optics (OOPAO) simulation tool \cite{heritierOOPAO}. We simulated an 8-meter telescope with a 41x41 DM and a PWFS, observing a natural guide star of magnitude 6.16. The time delay of the first stage was set to two frames. The atmospheric turbulence was a combination of nine frozen flow layers with Von Karman power spectra, with a Fried parameter of $15$ cm at $550$ nm wavelength. The simulation parameters can be found in Table \ref{table:simulator_parameters2}. We controlled the simulated system using an integrator and recorded the residual turbulence after DM correction. This exact set of residual turbulence phase-screens is then replayed by the SLM for all our GHOST experiments.


\begin{table*}[ht]
    \centering
    \caption{Simulations parameters. For full wind profiles of the simulated atmosphere, see the GitHub repository.}
    \label{table:simulator_parameters2}
\begin{tabular}{ |c| c| c|} 
 \hline
 \multicolumn{3}{|c|}{Numerically simulated first-stage} \\
 \hline
         Parameter  & Value  &  Units  \\
 \hline
 Telescope diameter   &  8  & m     \\
 Obstruction ratio    &  0  & percent         \\
 Sampling frequency (atm)  &  2000  & Hz        \\ 
 AO loop frequency (DM)   &  1000  & Hz        \\
 NGS magnitude &   6.16   &      $\cdots$       \\ 
 WFS wavelength & 0.79 &  \textmu  m  \\
 Actuators  & 41 & across the pupil \\
 PWFS modulation & 3 & \textlambda /D \\
 KL modes & 900 & modes \\
 Integrator gain & 0.5 & $\cdots$  \\

 \hline
 \multicolumn{3}{|c|}{GHOST (second-stage)} \\
 \hline
         Parameter  & Value  &  Units  \\
 \hline  
 Sampling frequency (simulation)  &  2000  & Hz        \\
 Sampling frequency (real-time) & 350 & Hz \\
 Actuators  & 24 & across the pupil \\
 PWFS modulation & 4 & $\lambda$/D \\
 WFS and Science Cam wavelength & 770 & nm \\
 Light source & 6 \& 187 & $10^3$ camera counts/ WFS frame \\
 \hline
 \multicolumn{3}{|c|}{Atmosphere parameters} \\
 \hline
 Fried parameter     &  15  & cm @ 500 nm     \\
 Number of layers    &  9  & $\cdots$       \\
 Effective wind speeds   &  34  & m/s      \\
 $L_0$ ($m$)      & 30   &   m        \\ 
  \hline  
\end{tabular}

\end{table*}

\subsection{Time delay experiment}\label{sec:delay}
The AO system delay budget encompasses various factors that contribute to the overall delay. Initially, a minimum 1-frame delay cannot be avoided due to two primary sources: 0.5 frames result from frame integrations performed by the WFS camera, and another 0.5 frames arise from the sample and hold operations carried out by the deformable mirror (DM). This 1-frame delay is a fundamental limitation and is encountered in systems such as GHOST as well.

When operating GHOST at a frame rate of 350 Hz, each frame corresponds to a duration of 2.86 milliseconds (ms). The camera readout time for a subwindow, which is approximately 70 microseconds (\textmu s), along with the time taken by the COSMIC pipeline (ranging from 100 to 150 \textmu s) and Python control (ranging from 600-800 \textmu s, discussed in Section \ref{sec:latency}) and the rapid settling of the deformable mirror (less than 20 \textmu s), are all relatively insignificant compared to the frame duration. Consequently, the total delay experienced is primarily comprised of the unavoidable 1-frame delay, along with additional ~0.3 frames for the remaining processes, resulting in a delay of around 1.3 frames. Therefore, an extra frame should be added to the GHOST system to replicate the scenario encountered in systems like SPHERE.

When the SPHERE camera operates at its maximum frame rate, determined by the camera readout time (around 1380 Hz), an additional full frame delay is incurred, bringing the total delay to 2 frames. Additionally, smaller overheads such as real-time control (RTC) computation, communication, and DM settling contribute to the overall delay, typically amounting to a few hundred microseconds. When all these factors are combined, the total delay experienced by the SPHERE system reaches approximately 2.4 frames \cite{10.1117/1.JATIS.2.2.025003}. However, if the system is run at a slower speed where the readout time is less than one frame, the total delay is reduced accordingly.

To study the method's predictive ability and demonstrate its ability to adapt to unknown time delays, we did the following test: We ran PO4AO (with the same hyperparameters) on three different temporal delays. We control these delays by adding an external buffer that suspends the DM commands for a given number of frames on the second stage only. First with zero frames of additional delay, second with one extra manually added frame of delay, and then with two (2) frames of additional delay. All test runs are compared against an integrator, which is separately tuned for all delays to give the best performance (lowest WFS residual variance). 

\begin{figure}
\begin{center}
\begin{tabular}{c}
\includegraphics[height=10.5cm]{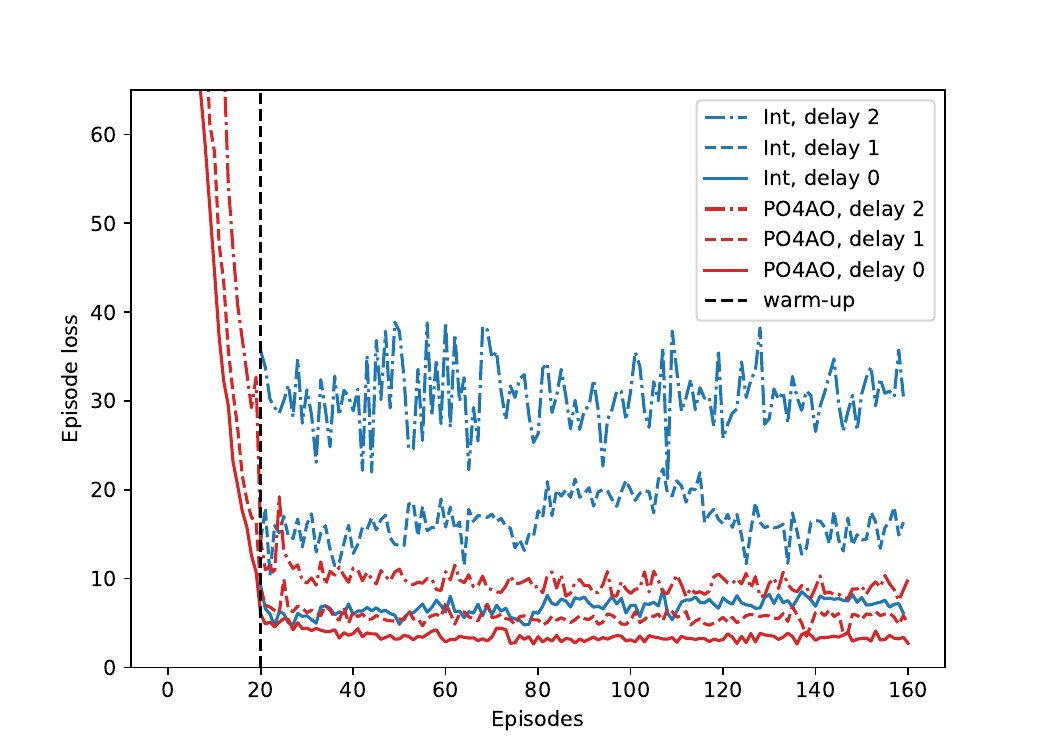}
\end{tabular}
\end{center}
\caption 
{ \label{fig:timedelay1}
Learning curves for time delay experiments. The red lines correspond to the performance of PO4AO during each episode, and the blue lines are for the integrator. A single episode is 500 frames. The gray dashed line marks the end of the integrator warm-up for PO4AO. In all cases, the PO4AO outperforms the integrator all ready after the warm-up period. The training is done parallel to control, so the 10 episodes correspond to approximately 14 sec in the figure.} 
\end{figure}

We compare the performance of PO4AO and the integrator in two ways: by comparing the reward, that is, the residual variance of the WFS measurements, and then by comparing the science camera images. the light source is set relatively bright, resulting in around 187k camera counts/frame with a standard deviation of 214. Figure \ref{fig:timedelay1} plots the learning curves, the cumulative wavefront residual variance (i.e., negative reward), after each episode during the run. The first 20 episodes are the warm-up phase, where the integrator controls the system with added Gaussian noise on the control signal; after each warm-up episode, the noise is reduced linearly. For time delays, PO4AO outperforms the integrator after the warm-up of 20 episodes (10k frames), and the performance converges at around 80 episodes (40k frames). Converged PO4AO provides around a factor of 3 improvement in reconstructed wavefront variance for all delays. 

\begin{figure}
\begin{center}
\begin{tabular}{c}
\includegraphics[height=9.5cm]{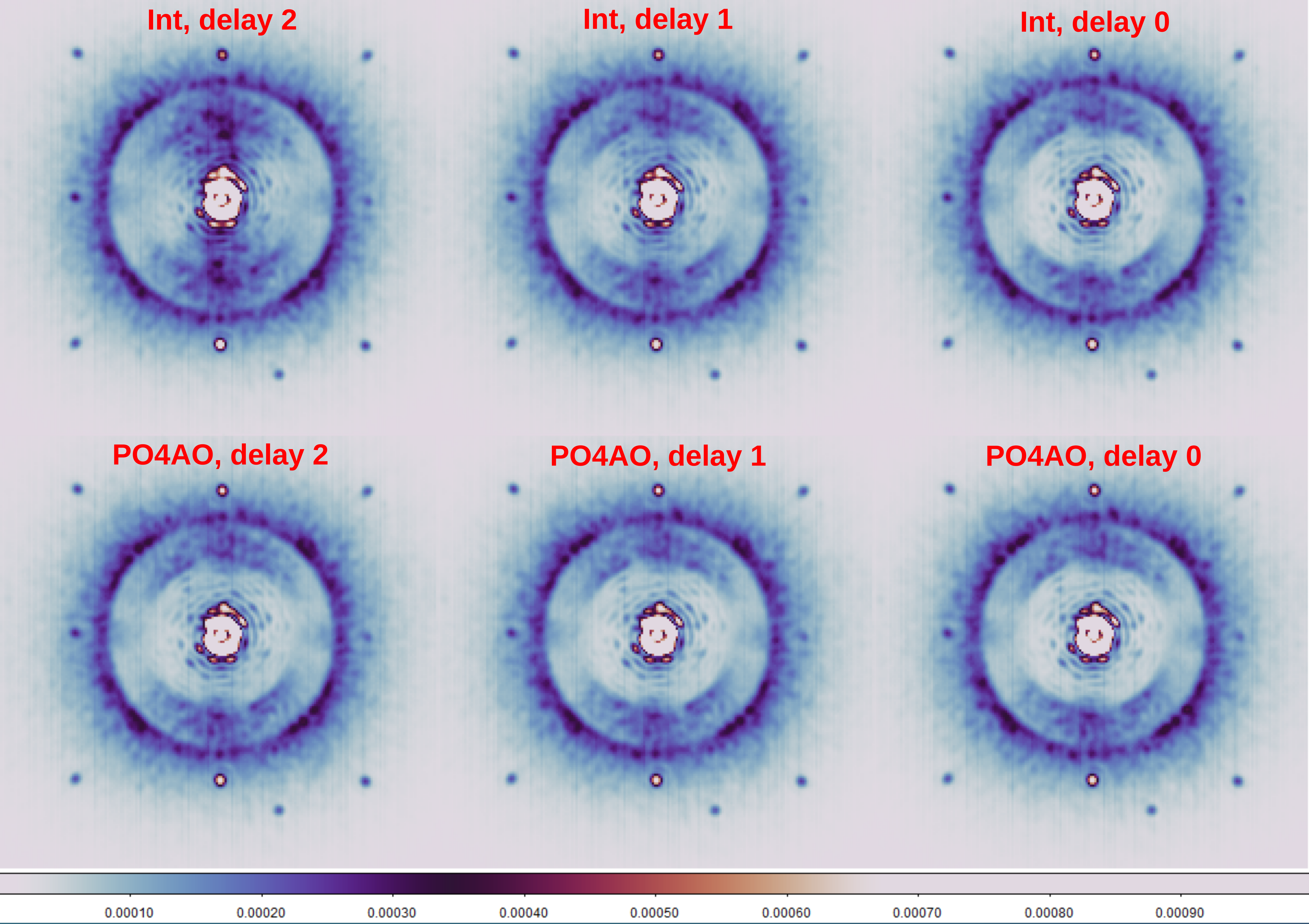}
\end{tabular}
\end{center}
\caption 
{ \label{fig:timedelay_psf}
PSFs on different additional control delays. The top row is for the integrator control, and the bottom is for PO4AO. The PO4AO and its hyper-parameters are exactly the same for all time delays -- the time delay is learned from the interaction.}
\end{figure}

\begin{figure}
\begin{center}
\begin{tabular}{c}
\includegraphics[trim={6cm 7cm 6cm 6cm},height=10.5cm]{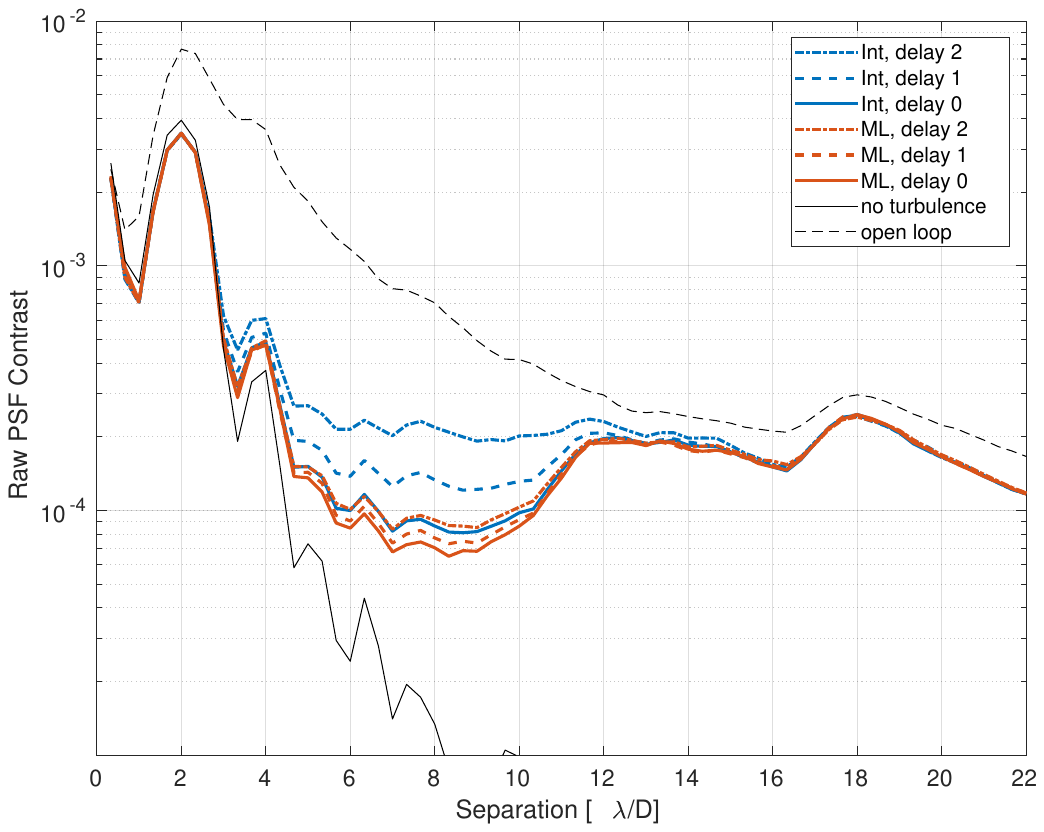}
\end{tabular}
\end{center}
\caption 
{ \label{fig:contrast}
The contrast with different time delays. The blue lines correspond to the azimuthal average of integrator images, and the red lines are the same for PO4AO. The line style indicates the length of the time delay. The solid black line is for flat DM, and the dashed black line is for flat DM (open loop) with first-stage residuals played on SLM. Note that around 11 \textlambda /D is the correction area of the DM-492, and around 18 \textlambda /D is the correction area of the numerically simulated 1st-stage DM. PO4A0 provides better contrast inside the 2nd-stage control radius for all time delays.} 
\end{figure}

Figure \ref{fig:timedelay_psf}  shows the long exposure (~8 sec) science camera images of the integrator and converged PO4AO after the first 80 episodes. Figure \ref{fig:contrast} plots the related azimuthal average over intensities of the images, that is, the contrast. Along with the PO4AO and the integrator results, we plot the contrast without turbulence which is limited by the bench non-common path aberrations (NCPA). We also plot the open-loop PSF contrast, where the SLM replays the numerically simulated 1st stage residuals.  We observe a factor of 1.5 - 3 improvement in contrast (depending on angular separation and time delay).

\subsection{Low flux experiment}
\begin{figure}
\begin{center}
\begin{tabular}{c}
\includegraphics[height=8cm]{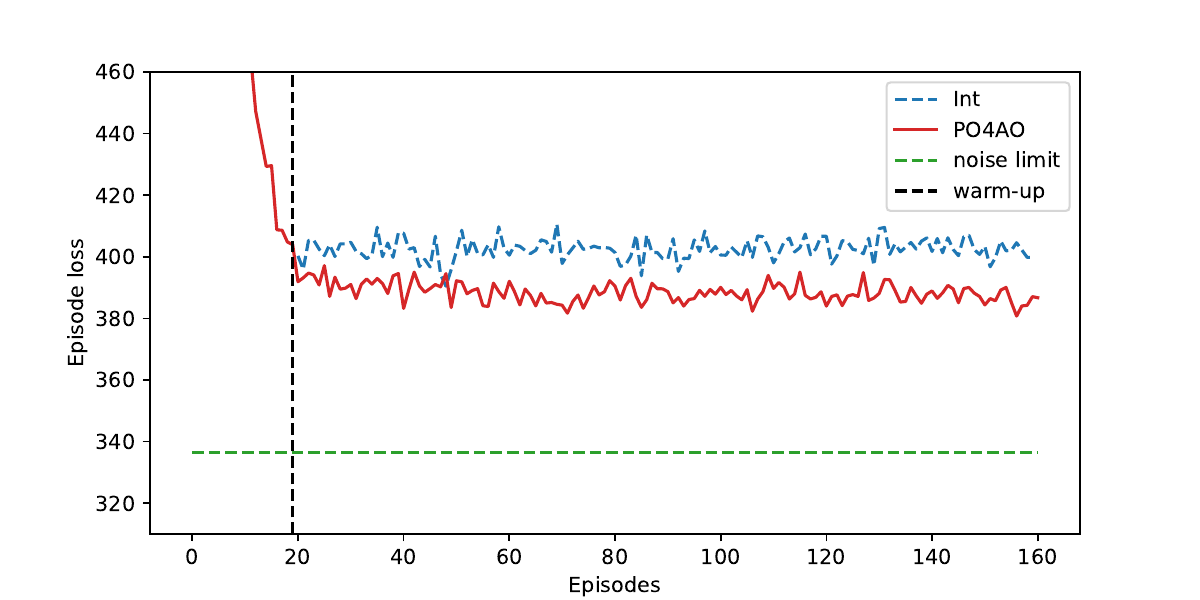}
\end{tabular}
\end{center}
\caption 
{ \label{fig:lowflux_train}
Learning curve for the low flux experiment. The blue line is the cumulative reward after each episode for the integrator, and the red line is for PO4AO. The dashed green line is the reward after each episode when the turbulence was not played and the loop opened. } 
\end{figure}

\begin{figure}
\begin{center}
\begin{tabular}{c}
\includegraphics[height=5.5cm]{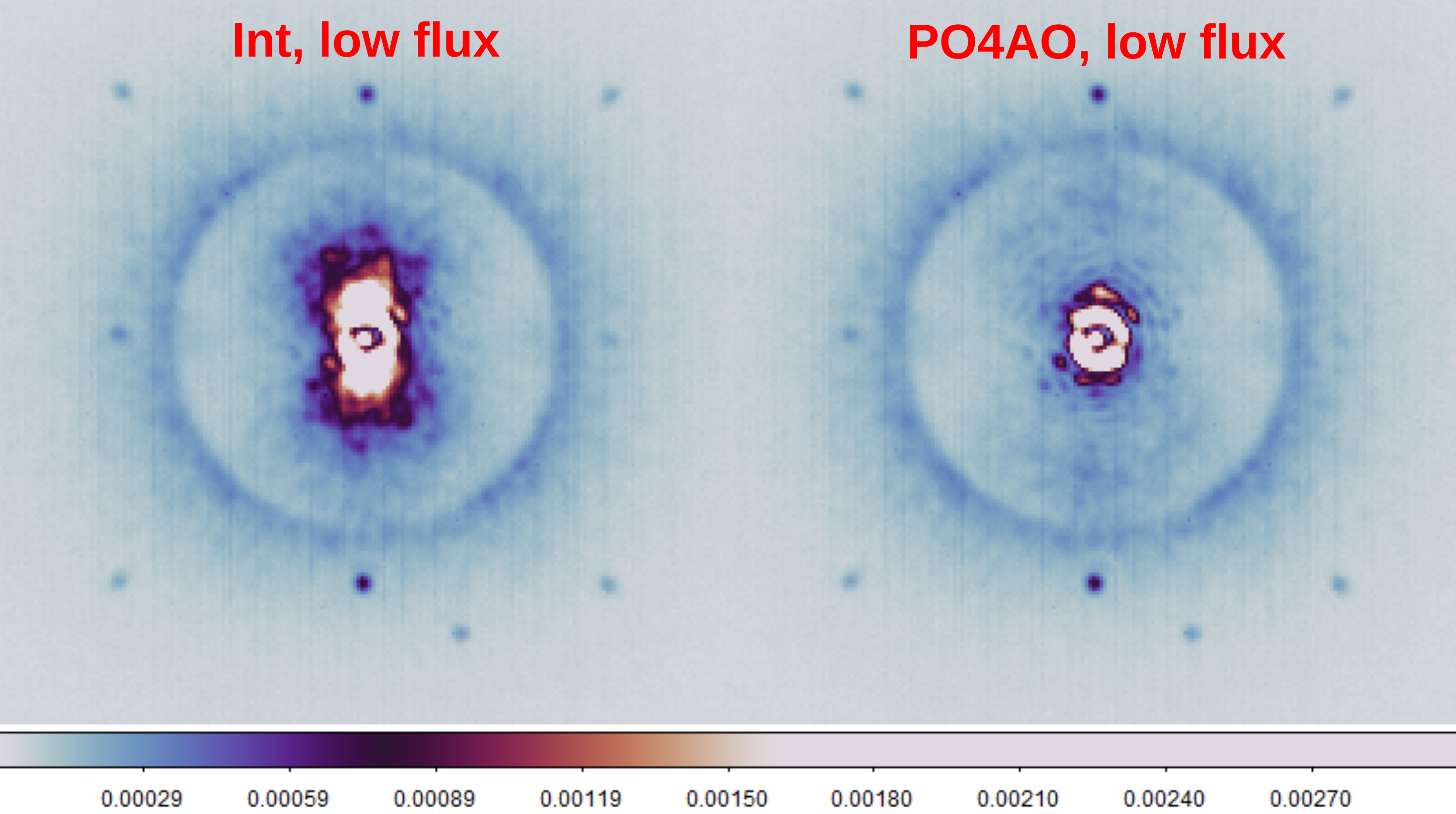}
\end{tabular}
\end{center}
\caption 
{ \label{fig:lowflux_psf}
PSFs during the low flux experiment. The left image is for the integrator, and the right is for the PO4AO. } 
\end{figure}

In this experiment, we study the performance of PO4AO under low flux. In the time-delay experiment, the internal light source was set relatively bright, resulting in a flux of 187k camera counts/frame. For the low flux experiment, we set the light source such that the flux was around 6000 camera counts/frame, while the detector noise was around 3000 camera counts/frame(S/N $\approx 2$). Further, we fixed the additional control delay to 1 frame, resulting in a typical overall delay of slightly over two frames, and ran the algorithm for 140 episodes, including the warm-up. Again, we record cumulative loss (negative reward) after each episode (Figure \ref{fig:lowflux_train}) and the science camera PSF after 80 episodes (Figure \ref{fig:lowflux_psf}). The cumulative loss here is dominated by photon and sensor noise; hence, we plotted the cumulative reward when turbulence is not played. It represents the absolute lower limit for the performance.  Figure \ref{fig:lowflux_psf} shows that PO4AO considerably reduces the photon flux inside the control radius. Consequently, PO4AO enables the imaging of fainter objects.

\subsection{Vibration reduction and effect of MDP formulation}\label{sec:horizon}
\begin{figure}
\begin{center}
\begin{tabular}{c}
\includegraphics[height=9.5cm]{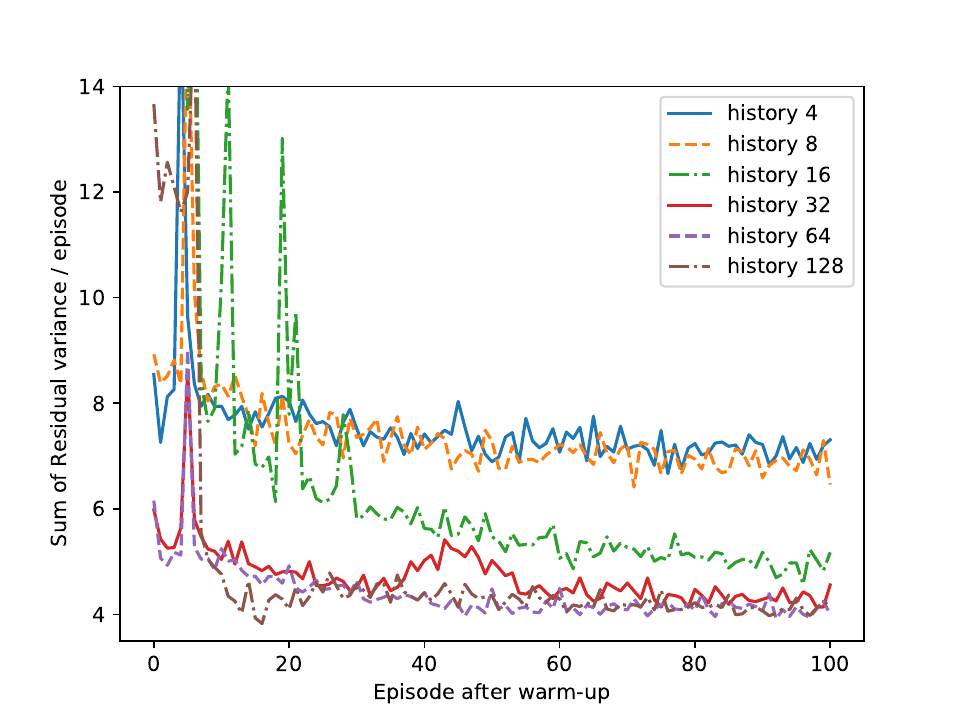}
\end{tabular}
\end{center}
\caption 
{ \label{fig:history}
Learning curves for different history lengths. We plot the cumulative reward after each episode for all history lengths after the warm-up phase. All history lengths have the same warm-up phase (not included in the plot). The longer the time length is, the better the performance.} 
\end{figure}

\begin{figure}
\begin{center}
\begin{tabular}{c}
\includegraphics[height=9.5cm]{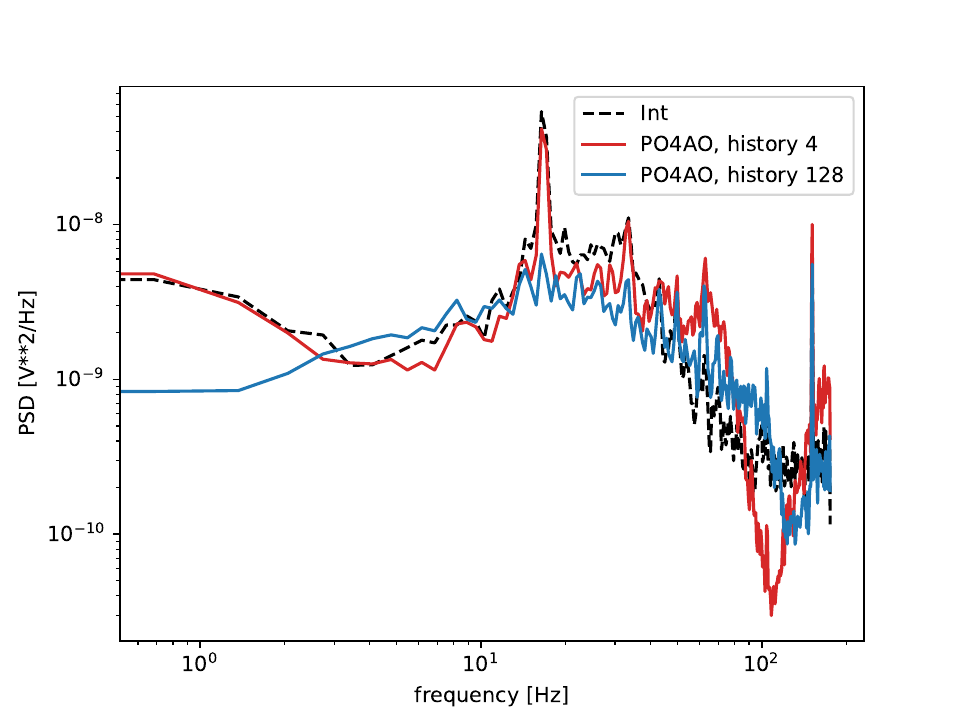}
\end{tabular}
\end{center}
\caption 
{ \label{fig:psds}
Temporal PSD of mode \# 1 (tip) on different history lengths} 
\end{figure}

This section studies the effect of the number of past telemetry frames (observations + actions) in the MDP formulation. We set the control delay to one extra frame and ran the same test but now with different history lengths. We note here that this test was run with a slower simulated wind profile (effective wind speed 26 m/s; the experiment could not be repeated with the original wind profile due to malfunctioning SLM); hence the lower rewards. Figure \ref{fig:history} shows the corresponding episode cumulative losses after the warm-up phase for different history lengths. As a general trend, the PO4AO corrections performance improves with the number of history frames considered. To better understand the increased predictive power with the number of frames, Figure \ref{fig:psds} shows the temporal power spectral densities of the mode \#1 for the history length of 128 frames, history length of 4 frames, and the integrator. The mode \#1 is tip mode in the direction of the dominant wind (the ground layer).

We observe that the PSD increases towards the Nyquist limit because of the intersampling signal of the 1st stage\cite{cerpa2022cascade}. The 1st stage is numerically simulated with a frame rate that is half the one in the 2nd stage; the 1st stage DM is only updated every second frame of the simulation. Hence, the location of the intersampling signal matches the Nyquist frequency of the 2nd stage and creates this particular shape of the PSD. 

Compared to the integrator, PO4AO dampens the residuals at mid-frequencies. However, like for linear controllers, we observe the Bode theorem-like behavior (waterbed effect); we observe amplification at higher frequencies. Some low-frequency residuals are transmitted through the leak. 


In addition, we observed a vibration spike at 16Hz. The spike is equally strong for the integrator, and PO4AO with four history frames, but the PO4AO with 128 history frames dampens the spike. Also, PO4AO with 32 and 64 could dampen the spike to some extent but not all the way as the PO4AO with 128 history frames. 

\subsection{Misregistration experiment}
Finally, we ran an experiment to confirm the method's robustness for Misregistration. We start with a pre-trained PO4AO and well-tuned integrator. The system is calibrated, the PO4AO is trained, and the integrator is tuned as before with no misregistration. Then we close the loop and start to introduce various degrees of misregistration. We introduce misregistration by shifting the DM off-axis manually while the loop is closed. 

\begin{figure}
\begin{center}
\begin{tabular}{c}
\includegraphics[height=7.5cm]{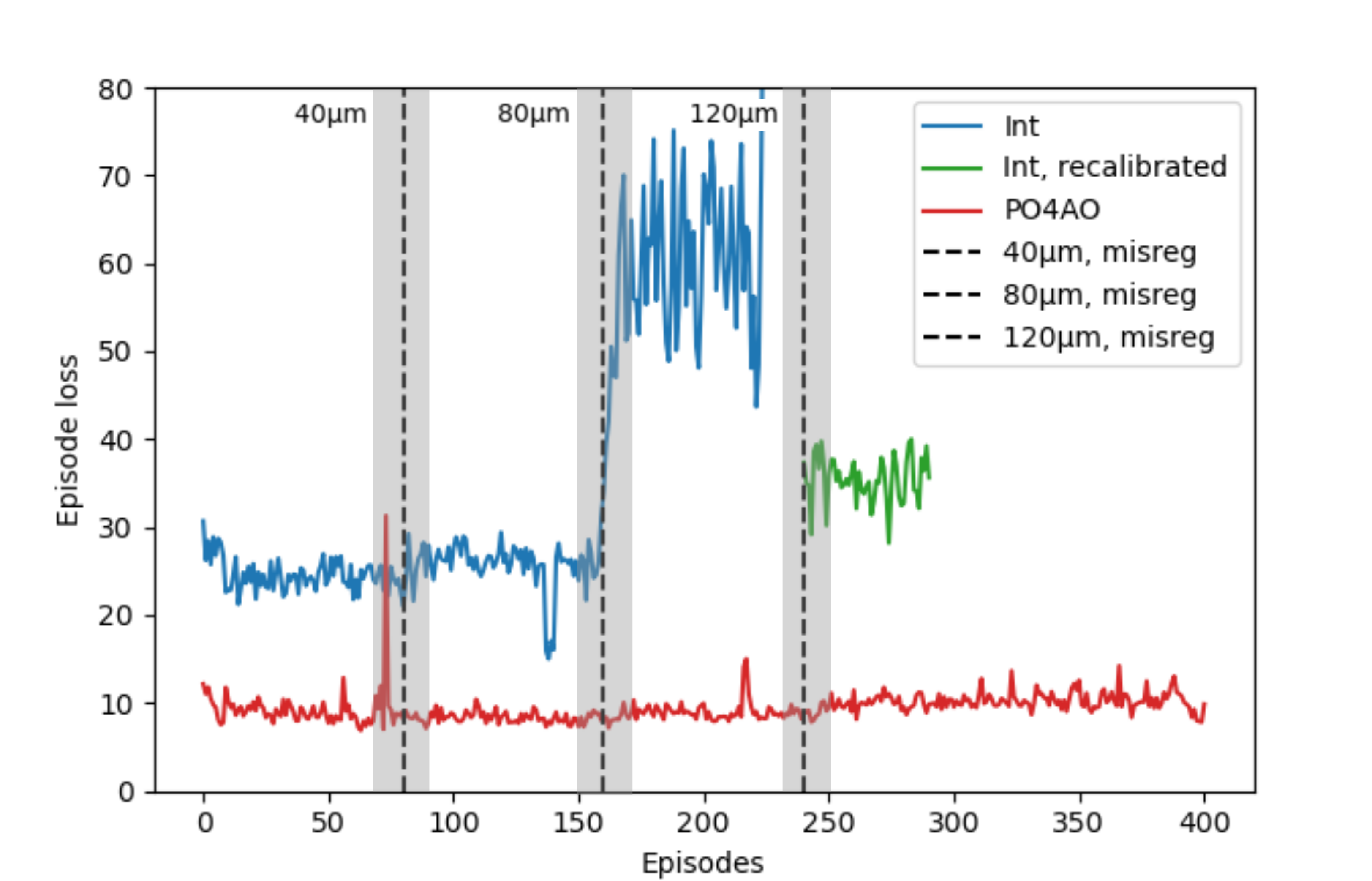}
\end{tabular}
\end{center}
\caption 
{ \label{fig:misreg}
The misregistration experiment. Here, we plot the cumulative loss over an episode during the misregistration experiment. The blue line is the well-tuned integrator calibrated with centered DM, and the green line is the well-tuned integrator calibrated with DM 120 microns off-axis. The red line is the PO4AO calibrated and pre-trained with centered DM. The dashed black lines indicate the moment when  DM was manually shifted. Since the shifting was done manually, the gray areas around the line indicate the uncertainty of the exact moment.  
} 
\end{figure}

The whole experiment lasts 320 episodes. The first 80 episodes are run with no misregistration. We then shift the DM by 40  micrometers (\textmu m) and let it run for 80 episodes; then at approximately 160 episodes, we shift the DM an additional 40 \textmu m (combining 80 \textmu m), and finally, at 240 episodes, we shift the mirror by 40 \textmu m (combing 120 \textmu m) and let it run 80 episodes. The DM actuator spacing is 300 \textmu m, meaning that 120 \textmu m corresponds to a 40\% shift. We repeat the same experiment with a well-tuned integrator. Finally, for reference, we re-calibrated (new interaction matrix and tuned gain) the system at 120 \textmu m of misregistration and ran a well-tuned integrator with a re-calibrated reconstruction matrix (the number of modes is kept the same). The results of this experiment are shown in Figure \ref{fig:misreg}. PO4AO is able to obtain stability and performance with dynamic misregistration while the integrator gets unstable. However, we observed PO4AO some instability when we first moved the mirror (see the spike around episode (76)), but PO4AO automatically recovered from this. PO4AO Also outperforms the re-calibrated integrator by a large margin.


\subsection{Performance on RTC}\label{sec:latency}
This section discusses the latency introduced by the method. The total delay budget of the COSMIC pipeline is discussed previously in Section \ref{sec:delay} and results in approximately 150 \textmu s (preprocessing of WFS data, projection to voltages, and clipping stage). In addition, there is a delay coming from the method's Python implementation. This includes all latency starting from when COSMIC writes the \emph{residual volts} to the shared memory to the point when the Python interface writes a set of new commands to the shared memory, and the loop resumes.   

More precisely, this latency is composed of the shared memory modification, the time required for converting vectors to images, and vice versa, as convolutional neural networks (CNNs) typically process images as input. Furthermore, the process of data collection adds to the overall delay, along with the delay caused by the update of the state memory of past telemetry frames. Finally, the delay budget also accounts for the time the policy neural network takes for its forward pass, which is essential for control decisions and accounts for the majority of the total latency. Table \ref{table:latency} shows the latency of different components for the most common CNN (32 history steps) used in the experiment. Collectively, these factors contribute to the total latency experienced in the system's operation. Table \ref{table:latency} shows that the most significant amount of time (500 \textmu s) is spent on the CNN forward pass. Saving data and updating the state information combines around 180 \textmu s. The remaining $\approx 200$ \textmu s is spent on image-to-vector modifications.

\begin{table*}      
\centering  
\caption{Latency terms of control thread.}  
\begin{tabular}{|c| c| c| c| } 
 \hline\hline
 \multicolumn{4}{|c|}{Inference speed} \\
 \hline
        & CNN inference \& jitter  & Saving data & update of the state     \\  
 \hline                                   
    CNN (32 history)  & 532.86+-8 \textmu s &  52.63 \textmu s & 128.38 \textmu s \\   
\hline
\end{tabular}
\label{table:latency}
\end{table*}

We further examine the latency by recording the end-to-end latency of CNN with different history lengths and PO4AO in the integrator mode (integrator running through the Python interface). By end-to-end latency, we mean the time from the WFS image arriving to the time the command is sent to the DM. Further, we measure the speed of the training procedure for each case given the training parameter (see Table \ref{table:simulator_parameters2}). The training parameters were chosen such that the training procedure fit inside a single episode for each history length. These results are shown in Table \ref{table:total_latency}. The integrator is around 500 \textmu s (time of single CNN forward pass) faster, as expected. The longer state history does not affect the overall loop latency. The longer state history only includes more convolutional filters (in the first layers). Hence, these introduce mostly parallel operations, and the GPU we used has plenty of memory headroom for the shallow architectures we used. However, the training time increases when more telemetry frames are added. This sets a limit to the episode length. Consequently, PO4AO, with fewer history frames, can adapt faster to misregistration and atmospheric wind conditions.

\begin{table*}            
\centering       
\caption{The total latency of Python implementation.  }  
\begin{tabular}{|c| c| c| c| c| } 
 \hline\hline
 \multicolumn{5}{|c|}{Total latency} \\
 \hline
          & Past frames (k \& m) & total latency & Jitter std.  & Tr. time / episode     \\  
 \hline                                   
    Integrator & - & 724 \textmu s & 85 & - \\
    CNN  &  4 &  1205 \textmu s & 60 & 0.78 sec  \\  
    CNN  &  8 & 1230 \textmu s  & 77 & 0.79 sec  \\ 
    CNN  &  16 & 1208 \textmu s  & 57 & 0.80 sec  \\ %
    CNN  &  32 & 1218 \textmu s  & 73 & 0.81 sec \\ 
    CNN  &  64 & 1219 \textmu s & 73 & 0.91 sec \\
    CNN  & 128 & 1196 \textmu s & 60 & 1.27 sec \\
\hline
\end{tabular}
\label{table:total_latency} 
\end{table*}

\section{Conclusion and discussion}
To conclude, this paper demonstrates that PO4AO is a robust controller for a second-stage AO system in a lab simulation. Reinforcement learning is shown to mitigate several critical error terms in XAO control, such as misregistration, photon noise, and temporal error. Moreover, Running PO4AO is a turnkey operation as the hyperparameters are tuned only when the method is implemented, and the method adapts automatically to changing conditions like noise level, misregistration, and wind profile. Extensive experiments on GHOST confirm that PO4AO can adapt to and mitigate these error terms on real hardware. In addition, we showed that PO4AO could also mitigate vibrations if it considers enough past telemetry frames.

However, like most deep RL methods, PO4AO is somewhat sensitive to the choice of hyperparameters. Tuning the parameters can take time, but the method performs robustly under all conditions once a good combination of hyperparameters is found. The method did not improve the system's stability to more degrees of freedom. We observed that the integrator was stable to approximately 350 KL modes, and PO4AO did not enable us to control more KL modes robustly. 

Additionally, we open-sourced the Python implementation of PO4AO, which is well-commented and can be easily adapted to numerical simulations or real hardware. The paper also discusses the hyperparameters of PO4AO and how they affect the method. The open-source implementation introduces an additional $\sim 700$ \textmu s to the pipeline latency, making it suitable for systems running lower than 1000Hz (depending on the pipeline latency) that can be tested on-sky with this implementation. The implementation requires a minimum of two GPUs on the RTC, one for inference and one for training. The memory bottleneck for bigger neural network structures (e.g., longer time length, more filters per layer, more DoF in the input) is the training GPU RAM; the backpropagation requires the most GPU memory.

Various avenues exist to optimize the method further: Firstly, critical code sections can be re-implemented by transitioning to lower-level programming languages such as C to benefit from direct memory access and optimized execution. Additionally, the NN could be implemented with NVIDIA TensorRT, which is a high-performance inference optimizer that delivers low latency and high throughput for deep learning inference applications. Secondly, optimizing memory handling through techniques like circular buffers can enhance data storage and retrieval efficiency, reducing computational load. Third, streamlining the pipeline to handle control voltage images instead of vectors can improve data processing and efficiency. Furthermore, the 500 \textmu s forward pass is slower than expected for the shallow architecture we used. Our investigation indicates a CPU bottleneck somewhere. Once the code is optimized, we expect significant gains in latency. These optimizations could provide latency gains that would support loop speed up to a few kilohertz.

\appendix    

\section{Implementation details}
\label{sect:misc}
This section discusses open-source Python implementation. The code consists of three Python files: \path{po4ao.py}, \path{po4ao_models.py},  \path{po4ao_util.py}, and \path{po4ao_config.py}. The \path{po4ao.py} script includes the main function that starts the warm-up phase and training procedure and closes the loop. The file includes the \emph{step}-function that interfaces the Python code to the RTC pipeline. The PO4AO algorithm interfaces to the COSMIC pipeline through a so-called step-function. The step function can be found in the \path{po4ao.py} file. The step function is pipeline-specific and should be replaced if the method is adapted to different RTC pipelines or numeric simulations. The \path{po4ao.py} file also includes functions for the training thread, control thread, and warm-up phase. The file \path{po4ao_models.py} contains the NN models: the policy and the dynamics, and file \path{po4ao_util.py} includes utility functions, such as the implementation for Replay buffers and shared memory optimizers for training. The configuration file \path{po4ao_config.py} includes all adjustable hyperparameters of the method.

The file \path{po4ao_config.py} contains few parameters beyond the PO4AO parameters discussed in Section \ref{sec:imp}. In the following, give a short explanation of them.

\subsection{Parameters under subcategory the integrator}
If the parameter \emph{integrator} is set to True, the PO4AO runs in integrator mode, where data is collected and models are trained, but the control thread runs on integrator control. These parameters define the internal integrator controller of POAO and affect the PO4AO itself. The gain value gives the gain of the integrator during either the warm-up phase or in the integrator mode. The "leakages" and "number of modes" set the leakages and the DoF of the internal integrator and PO4AO. The number of modes should be set to a value where the integrator is stable on a reasonable gain value.

\subsection{Parameters under subcategory the Neural network models}
The NN parameters are for the architecture of the Policy and the dynamics NNs. The models are generic 3-layer fully convolutional NN without any up- or down-sampling and share the same architecture excluding the output layer \cite{nousiainen2022toward}. The \emph{CNN filters/layer} sets the number of convolutional filters on layers. Different NNs architectures can be easily implemented in the code by replacing the models in the \path{po4ao_models.py}- file.

\subsection{Other parameters}
The control delay parameter (under the category "MDP parameters" in the code) sets the additional control delay. It decides how many frames a control signal is suspended on top of the natural loop latency. It is only for testing the behavior of PO4AO on different time delays. Obviously, for a real system, it is set to zero for optimized performance.


\subsection* {Acknowledgments}
The work of J.N. and T.H. was supported by the Academy of Finland (decisions 326961, 345720, and 353094). Part of this work was supported by an ETH Zurich Research Grant. S.P.Q. gratefully acknowledges the financial support from ETH Zurich. Part of this work has been carried out within the framework of the National Centre of Competence in Research PlanetS supported by the Swiss National Science Foundation (SNSF) under grants 51NF40 182901 and 51NF40 205606. S.P.Q. and A.M.G acknowledge the support from the SNSF. The COSMIC is developed through a strategic partnership between LESIA at Observatoire de Paris, AITC at the Australian National University, Microgate, and CAS at Swinburne University of Technology.

\subsection* {Code, Data, and Materials Availability} 
The codes used in this paper are available on GitHub repository [\href{https://github.com/jnousi/PO4AO.git}{https://github.com/jnousi/PO4AO.git}]. The code is documented and annotated to help readers understand the methodology and reproduce the results.

We encourage readers to use the data and codes for their own research and to cite this paper as the source of the data. If you have any questions about the data or the codes, please do not hesitate to contact us.


\bibliography{report}   
\bibliographystyle{spiejour}   


\vspace{2ex}\noindent\textbf{Jalo Nousiainen} is a postdoctoral researcher at LUT University. His research focuses on algorithm development for high-contrast instruments dedicated to exoplanet imaging. He has a background in applied mathematics, specifically in Inverse problems, Bayesian statistics, and machine learning.

\vspace{2ex}\noindent\textbf{Byron Engler} is a postdoctoral fellow at ESO, working in the field of adaptive optics. 

\vspace{2ex}\noindent\textbf{Markus Kasper} received his Ph.D. degree from the University of Heidelberg in 1999, after which he joined ESO to work on Adaptive Optics, High-Contrast Imaging, and astronomical instrumentation projects. He was involved in many of ESO's AO projects, such as MACAO and NACO LGS, and was ESO's project leader for VLT-SPHERE. He was the principal investigator of the ELT/EPICS phase-A study (2008-2010, predecessor of PCS) and of NEAR, the mid-IR imaging experiment to search for habitable planets in Alpha Centauri in collaboration with the Breakthrough Initiatives. He is currently working towards kicking off the ELT-PCS planet imager project.

\vspace{1ex}
\noindent Biographies and photographs of the other authors are not available.

\listoffigures
\listoftables

\end{spacing}
\end{document}